\documentclass[prd,preprint,nofootinbib,showpacs,preprintnumbers,amsmath,amssymb,eqsecnum]{revtex4-1}

\usepackage{graphicx}
\usepackage{bm}
\usepackage{subfigure}

\begin{document}

\preprint{RUP-14-11}

\title{Black holes as particle accelerators: a brief review}

\author{$^{1}$Tomohiro Harada}%
\email{harada@rikkyo.ac.jp}
\affiliation{$^{1}$Department of Physics, Rikkyo University, Toshima,
Tokyo 171-8501, Japan}
\author{$^{2}$Masashi Kimura }
\email{M.Kimura@damtp.cam.ac.uk}
\affiliation{$^{2}$Department of Applied Mathematics and Theoretical Physics, 
Centre for Mathematical Sciences, 
University of Cambridge, Wilberforce Road, Cambridge CB3 0WA, UK}
\date{\today}

\begin{abstract}
Rapidly rotating Kerr black holes can accelerate particles to
 arbitrarily high energy if the angular momentum of the particle 
is fine-tuned to some critical value.
This phenomenon is robust as it is founded on the basic 
 properties of geodesic orbits around a near-extremal Kerr black hole. 
On the other hand, the maximum energy of the acceleration is subjected
 to several physical effects.
There is convincing evidence that the particle acceleration to
 arbitrarily high energy 
is one of the universal properties of general 
near-extremal black holes.
We also discuss gravitational particle acceleration 
in more general context.
This article is intended to provide a pedagogical introduction 
to and a brief overview of this topic for non-specialists.
\end{abstract}


\maketitle

\tableofcontents

\newpage

\section{Introduction}
The existence of black holes is strongly 
suggested by astrophysical observations.
However, it is not so clear whether the observed objects 
are really identical with what we know as black holes 
predicted in general relativity.
It is necessary to understand the physics of 
black hole horizons for 
the direct observational confirmation of 
black holes. 
In this article, we will discuss the possibility of black holes
being particle accelerators. 
What do black holes as particle accelerators mean?

Before asking this question, let us ask what it means that 
terrestrial particle accelerators, such as the Large Hadron Collider,  
accelerate particles.
In most of terrestrial particle accelerators, the kinetic energy of 
particles is increased through the work exerted on the 
particles by electromagnetic force. 
For the proton-proton collision in the Large Hadron Collider,  
the energy $E_{\rm cm}$ measured by an observer who is at rest with respect to 
the centre of mass frame, which is called centre-of-mass (CM) energy
and denoted as $E_{\rm cm}$, 
becomes as high as 14 tera electron volts (TeV).
This is 15,000 times the rest mass energy of 
proton approximately.

Black holes have gravitational force strong enough to trap
light rays. The boundary behind which no light ray can escape to infinity
is called an event horizon. Since gravitational force acts
on both charged and neutral particles, 
black holes can accelerate not only charged particles but also 
neutral particles. From such a consideration, 
it can be regarded as natural that black holes accelerate 
particles. However, for the Schwarzschild black hole, which is a 
static spherically symmetric black hole, the CM energy 
$E_{\rm cm}$ of two particles of equal rest mass $m$ 
which have been at rest at infinity can be $2\sqrt{5}mc^{2}$
at most, which corresponds to $\gamma=9$ in terms of the 
relative velocity, and hence it cannot be regarded as 
a high energy particle accelerator. 

In 2009, Ba\~nados, Silk and West~\cite{Banados:2009pr}
found that $E_{\rm cm}$ can be
arbitrarily high for rotating black holes
in the context of dark matter particle annihilation at the 
galactic centre. It should be noted that the unboundedly high $E_{\rm
cm}$ of particle collision had already been 
noticed in a different context in Ref.~\cite{Piran_etal_1975}.
Because of 
the equivalence principle of general relativity, 
not only microscopic particles such as electrons, 
protons, neutrons, ions and molecules but also 
macroscopic objects such as black holes and 
compact stars can be accelerated by a rotating black hole with the same 
gamma factor if the size and mass of those objects are sufficiently
small compared to the those of the central black hole.

This paper is organised as follows. In Section II, we elucidate 
that the CM energy of two colliding particles on the equatorial 
plane of a Kerr black hole can be arbitrarily high and discuss why 
this is possible in contrast to the case of a Schwarzschild black hole.
In Section III, we review critical comments and basic questions 
on this scenario and respond to them from a physical point of view.
In Section IV, we discuss the possibility of particle acceleration in  
astrophysical black holes.
In Section V, we briefly review the generalisations of the particle acceleration
scenario.
Section VI is devoted to conclusion.
In the following, we use the unit where $G=c=1$ unless explicitly noticed.

\section{Kerr black holes as particle accelerators}

\subsection{Kerr black holes and geodesic particles}

In general relativity, 
a stationary rotating vacuum black hole is uniquely described by a Kerr spacetime.
The line element $ds^2 = g_{\mu \nu}dx^\mu dx^\nu$ 
in the Kerr spacetime is written in the Boyer-Lindquist
coordinates in the following form~\cite{Kerr1963,Wald:1984rg}:
\begin{eqnarray}
ds^{2}&=&-\left(1-\frac{2Mr}{\rho^{2}}\right)dt^{2}
-\frac{4Mar\sin^{2}\theta}{\rho^{2}}d\phi dt
+\frac{\rho^{2}}{\Delta}dr^{2}+\rho^{2}d\theta^{2} \nonumber \\
&&+\left(r^{2}+a^{2}+\frac{2Mra^{2}\sin^{2}\theta}{\rho^{2}}\right)
\sin^{2}\theta d\phi^{2},
\end{eqnarray}
where $\rho^{2}=\rho^{2}(r,\theta)=r^{2}+a^{2}\cos^{2}\theta$ and $\Delta=\Delta(r)=r^{2}+a^{2}-2Mr$.
Kerr black holes are parametrised by 
mass $M$ and spin $a$,
which must satisfy $0\le |a| \le M$~\cite{Poisson2004}.
The spin parameter $a$ is related to
the angular momentum with respect to the rotational 
axis of the black hole as $J = M a$.
Black holes with the maximum value of the spin parameter are called 
maximally rotating black holes. 
We assume $a\ge 0$ without loss of generality. For later use, we 
define the non-dimensional spin parameter $a_{*}=a/M$.

The Kerr spacetime is stationary and axisymmetric with corresponding 
Killing vectors $\partial/\partial t$ and 
$\partial/\partial \phi$.
The event horizon is located at 
$r=r_{H}=M+\sqrt{M^{2}-a^{2}}$, where $\Delta(r)$
vanishes, and is rotating with the angular velocity $\Omega_{H}=a/(r_{H}^{2}+a^{2})$.
Note that the horizon is called extremal when
$r = r_H$ is a double root of $\Delta(r)$, i.e., $a = M$~\footnote{For a general stationary black hole,
an extremal horizon is defined as a horizon 
on which the surface gravity is zero~\cite{Poisson2004}.
}.
The region given by $r_{H}<r<
r_{E}(\theta)=M+\sqrt{M^{2}-a^{2}\cos^{2}\theta}$, 
where the Killing vector $\partial /\partial t$ of stationarity
is spacelike, is called an ergoregion.

In general relativity, a free test particle moves along a geodesic of the 
spacetime. The energy $E=-g_{t\mu}p^{\mu}$ 
and angular momentum $L=g_{\phi \mu}p^{\mu}$ 
of the particle with four-momentum $p^{\mu}$
are conserved in accordance with the symmetries of the spacetime,
where we can write $p^{\mu}=\dot{x}^{\mu}$ with the dot being the 
derivative with respect to the affine parameter $\lambda$.
It should be noted that 
this energy $E$ is with respect to an observer at rest at infinity 
and is distinct from the CM energy $E_{\rm cm}$. 
Additionally, it is known that there is another conserved quantity
called the Carter constant, which is related to the total angular momentum. 
Because of the existence of the 
conserved quantities $E$, $L$ and the Carter constant,
geodesic equations are integrated to first-order differential
equations.

For a particle which moves on the equatorial plane, for which 
the Carter constant vanishes, 
$\dot{t}$ and $\dot{\phi}$ can be written as  
\begin{equation}
r^{2}\dot{t} = \frac{(r^2 + a^2)[E (r^2 + a^2) - aL] }{\Delta} 
- a(a E -L),
\label{dtdlambda}
\end{equation}
and 
\begin{equation}
r^{2}\dot{\phi} = \frac{a[E (r^2 + a^2) - aL] }{\Delta} 
- (a E -L),
\label{dphidlambda} 
\end{equation}
respectively.
Using $p^{\mu}p_{\mu}=-m^{2}$, 
the geodesic equation for a particle 
on the equatorial plane is reduced to a simple 
one-dimensional potential problem given by 
\begin{equation}
\frac{1}{2}\dot{r}^{2}+V(r)=0,
\label{energyequation}
\end{equation}
with
\begin{eqnarray}
V(r)&=&-\frac{[(r^{2}+a^{2})E-aL]^{2}-\Delta(r)[m^{2}r^{2}+(L-aE)^{2}]}
{2r^{4}} \nonumber \\
&=&-\frac{m^{2}M}{r}+\frac{L^{2}-a^{2}(E^{2}-m^{2})}{2r^{2}}-\frac{M(L-aE)^{2}}{r^{3}}-\frac{1}{2}(E^{2}-m^{2}),
\label{eq:effective_potential}
\end{eqnarray}
where $m$ and $V(r)$ are called the rest mass and the effective
potential of the particle, respectively.
For a massive particle, we have $p^{\mu}=m u^{\mu}$, 
where $u^{\mu}$ is the four-velocity satisfying $u^{\mu}u_{\mu}=-1$.
The motion is possible only in the region where $V\le 0$. 

Physical particles must satisfy the so-called forward-in-time condition
which guarantees that the value of the time coordinate $t$ increases 
along the trajectory of the particle's motion.
Near the event horizon,
Eq. (\ref{dtdlambda}) implies that this condition reduces 
to $E-\Omega_{H}L\ge 0$. 
Here we call 
particles which satisfy $E-\Omega_{H}L=0$ critical particles.

\subsection{Particle collision in the equatorial plane}

If two particles 1 and 2 are at the same spacetime
point, an observer at the centre-of-mass frame is defined as 
the one whose four-velocity is parallel to 
the sum of the four-momenta $p_{1}^{\mu}$ and $p_{2}^{\mu}$
of particles 1 and 2, respectively. The CM energy  
$E_{\rm cm}$ is defined as the energy measured by this observer 
and is given by 
\begin{equation}
E_{\rm
 cm}^{2}=-g_{\mu\nu}(p_{1}^{\mu}+p_{2}^{\mu})(p_{1}^{\nu}+p_{2}^{\nu}).
\label{eq:Ecm}
\end{equation}
The CM energy is scalar invariant and physically observable in principle. 

\begin{figure}[tbh]
\begin{center}
 \includegraphics[width=0.3\linewidth]{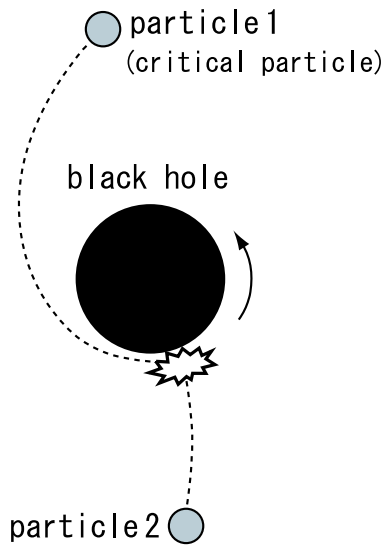}
\end{center}
\caption{
The schematic figure of particle collision 
for which the CM energy can be very large. 
The arrow denotes the direction of 
the spin of the black hole. 
For arbitrarily high CM energy to be achieved, 
the critical particle orbits the black hole 
arbitrarily large number of times 
with arbitrarily long proper time.
\label{BSW_critical_picture}}
\end{figure}

Let us first concentrate ourselves on two 
particles both of which move in the equatorial plane
and collide with each other near the horizon.
Figure~\ref{BSW_critical_picture} schematically illustrates the situation 
under consideration.
In this case, the CM energy $E_{\rm cm}$ for the collision 
in the vicinity of the horizon is calculated to 
give~\cite{Harada:2010yv,Harada:2011xz}
\begin{eqnarray}
E_{\rm cm}^{2}=
\frac{m_{1}^{2} r_{H}^{2}+(L_{1}-aE_{1})^{2}}{r_{H}^{2}}
\frac{E_{2}-\Omega_{H}L_{2}}{E_{1}-\Omega_{H}L_{1}}
+(1\leftrightarrow 2)
+\cdots,
\label{eq:Ecm_equatorial}
\end{eqnarray}
where two labels 1 and 2 denote particles 1 and 2, respectively, 
``$(1 \leftrightarrow 2)$''  denotes the term which is obtained by 
exchanging 1 and 2 in the first term, and ``$\cdots$'' denotes the 
terms which are obviously finite. Therefore, $E_{\rm cm}$ is divergent 
if either of particles 1 and 2 satisfies the critical condition
$E - \Omega_H L =0$.
It should be noted that $E_{\rm cm}$ is shown to be bounded if the
collision point is not near the horizon.

\begin{figure}[tbh]
\begin{center}
\includegraphics[width=0.6\linewidth]
{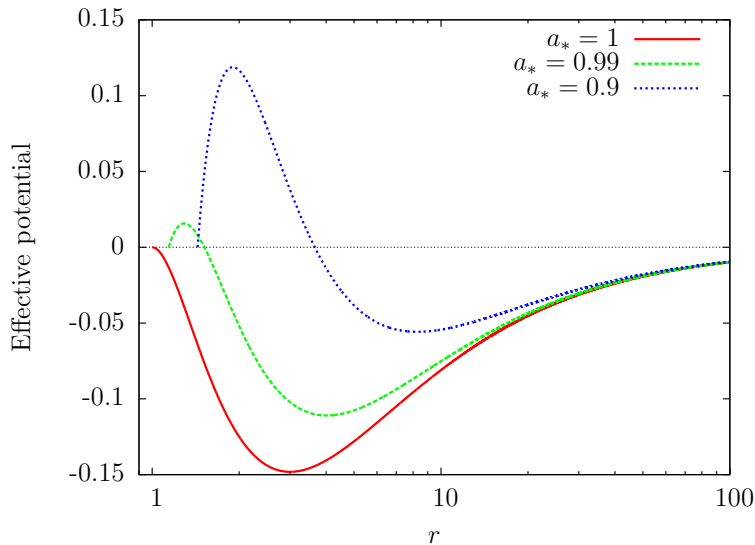}
\caption{\label{fg:potential}
The effective potential defined by Eq.~(\ref{eq:effective_potential})
for marginally bound critical particles, where $r$ and $V$ are
 normalised by $M$ and $m^{2}$, respectively.
$V$ must be non-positive in the allowed region of particle motion.
}
\end{center}
\end{figure}

For such a high energy collision to occur, the critical particle has to
approach the horizon as a result of its motion. 
For a while, 
we concentrate ourselves on massive particles of rest mass 
$m$ in the equatorial plane. 
It is convenient to define non-dimensional specific energy
$e=E/m$ and specific angular momentum $l=L/(mM)$.
We introduce $l_{c}=E/(\Omega_{H}mM)$ as 
the critical value for $l$.
The motion of geodesic particles on the equatorial plane 
can be fully analysed by the 
effective potential given by Eq.~(\ref{eq:effective_potential}) 
and is extremely simple for a particle 
which is initially at rest at infinity, for which $e=1$.
In this case, we find 
\begin{equation}
 V(r_{H})=-\frac{(r_{H}^{2}+a^{2})^{2}(m-\Omega_{H}L)^{2}}{2r_{H}^{4}}\le 0.
\end{equation}
If and only if $V(r)< 0$ for $r_{H}<r<\infty$, or equivalently, 
the quadratic equation
\begin{equation}
 2 r^{2}-Ml^{2}r+2M^{2}(l-a_{*})^{2}=0
\end{equation}
has no root in the region $0<r<r_{H}$,
such a particle approaches the horizon from infinity.
This condition reduces to 
\[
-2(1+\sqrt{1+a_{*}})=l_{L}<l<l_{R}=2(1+\sqrt{1-a_{*}})
\].
Noting that $l_{R} \le l_{c}$ and the equality 
holds only for $a_{*}=1$, it turns out that 
the critical particle can 
reach the horizon for $a_{*}=1$
but cannot for $a_{*}<1$.
Figure~\ref{fg:potential} shows the effective potentials of critical particles
for $a_{*}=0.9, 0.99$ and $1$. We can see that the critical 
particle can reach the horizon only for $a_{*}=1$
with infinitely long proper time.
In the case of $a_* < 1$, if the spacetime is 
 near-extremal $a_* \simeq 1$,
 the critical particle reaches the radius 
\[
 r \simeq r_H + 
 \frac{2\sqrt{2}(E^2+m^2)M}{3E^2 -m^2}\sqrt{1-a_{*}}
\]
and is bounced back there.

Let us consider the collision between two particles 
of same rest mass $m$ which are 
initially at rest at infinity. If the
two particles collide in the vicinity of the 
horizon for $a_{*}=1$, Eq.~(\ref{eq:Ecm_equatorial}) gives~\cite{Banados:2009pr}
\begin{equation}
\frac{E_{\rm cm}}{2m}=\sqrt{\frac{1}{2}\left(\frac{2-l_{1}}{2-l_{2}}+\frac{2-l_{2}}{2-l_{1}}\right)}.
\end{equation}
Therefore, if we fine-tune either $l_{1}$ or $l_{2}$
to the upper limit angular momentum $l_{R}=2 (= l_c)$, 
$E_{\rm cm}$ can be arbitrarily large. 
If we choose $l_{1}=l_{R}$ and $l_{L}<l_{2}<l_{R}$ and 
set the collision point at 
$r_{\rm col} \simeq r_{H}=M$ for $a_{*}=1$, 
Eq.~(\ref{eq:Ecm}) gives~\cite{Harada:2010yv}
\begin{eqnarray}
\frac{E_{\rm cm}}{2 m} &\simeq & \sqrt{\frac{(2 - \sqrt{2})(2-l_{2})M}{2(r_{\rm col} - M)}},
\label{ecmnearhorizon}
\end{eqnarray}
which again diverges in the limit $r_{\rm col} \to r_{H}=M$.
On the other hand, if
$l_{1}=l_{R}(<l_{c})$ and $l_{2}<l_{R}$ for $a_{*}<1$,
Eq.~(\ref{eq:Ecm_equatorial}) gives~\cite{Jacobson:2009zg,Harada:2010yv}
\begin{equation}
\frac{E_{\rm cm}}{2m}\simeq \frac{\sqrt{(2+\sqrt{2})(2-l_{2})}}{2}
\frac{1}{\sqrt[4]{1-a_{*}^{2}}}
\end{equation}
for the collision in the vicinity of the 
horizon, where 
$a_{*}\simeq 1$ is assumed. 
We find $E_{\rm cm}\to \infty$ as $a_{*}\to 1$.

\subsection{CM energy in finite acceleration time}
\label{subsec:acceleration_time}
It would be 
physically important how high the 
CM energy for the collision can be
near the Kerr black hole if the acceleration continues for 
finite time. That is, 
we would like to estimate the maximum CM energy 
which can be achieved within finite time~\cite{Patil:timescale}.

Let us consider the collision of two particles 1 and 2 of same rest mass
$m$ which are at rest at infinity near a maximally rotating black hole.
We choose $l_1=l_{R}=2$ and $l_{L}<l_{2}<l_{R}=2$. 
Equations (\ref{dtdlambda}) and (\ref{energyequation}) implies
\begin{equation}
 \frac{dr}{dt}=\frac{(r-M)^{2}\sqrt{2M}}{(r^{2}+Mr+2M^{2})\sqrt{r}},
\label{eq:drdt_critical_extremal}
\end{equation}
so that it takes infinite Killing time $t$ 
for particle 1 to reach the horizon. 
Using Eq. (\ref{eq:drdt_critical_extremal}), 
we can calculate the Killing time $T$ needed for particle 1 
to reach the collision point $r = r_{\rm col}$ from 
a distant location $r = r_i $ around a maximally rotating Kerr black hole as
\begin{eqnarray}
T = -  \int_{r_i}^{r_{\rm col}}
d r 
\frac{\sqrt{r} (r^2 + M r  + 2M^{2} )
}{\sqrt{2M}(r-M)^2} \simeq \frac{2\sqrt{2}M^{2}}{r_{\rm col}-M},
\label{timescale1}
\end{eqnarray}
where we have assumed $M \ll r_i \ll M^{2}/(r_{\rm col} - M)$ in the
approximation on the right-hand side. 
This assumption is valid in reasonable astrophysical 
situations.
{}From Eqs. (\ref{ecmnearhorizon}) and (\ref{timescale1}), 
we obtain
\begin{equation}
\frac{E_{\rm cm}}{2m} \simeq
 \frac{1}{2}\sqrt{(\sqrt{2}-1)(2-l_{2})\frac{T}{M}},
\end{equation}
or 
\begin{equation}
E_{\rm cm}\simeq
2.5 \times 10^{20} {\rm eV}
\left(\frac{T}{10~{\rm Gyr}}\right)^{1/2}
\left(\frac{M}{M_{\odot}}\right)^{-1/2}
\left(\frac{m}{1~{\rm GeV}}\right). 
\end{equation}
It is interesting to note that the maximum CM energy which can be
achieved within the age of the universe is as high as ultra high energy
cosmic rays.

\subsection{Physical explanation of particle acceleration}

\begin{figure}[tbh]
\begin{center}
\includegraphics[width=0.45\linewidth]{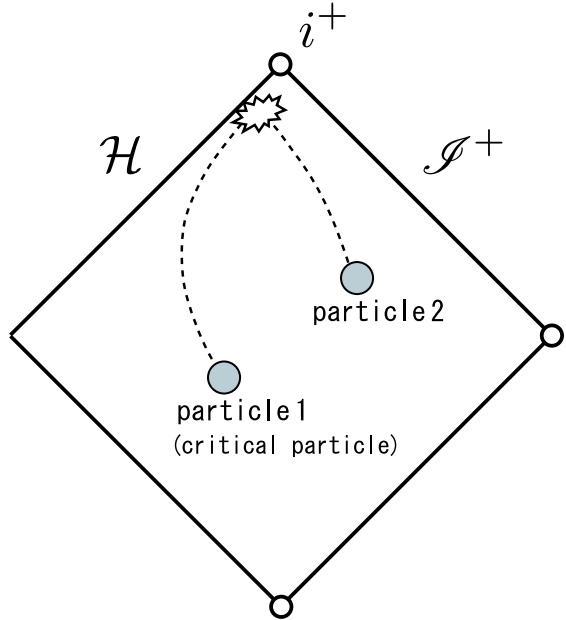}
\end{center}
\caption{\label{fg:conformal_diagram}
The spacetime diagram of the high energy particle collision between 
critical and non-critical particles near the event horizon ${\cal H}$.
The critical particle asymptotes the event horizon in infinite proper time.
Since the only possible causal curve on the event horizon is light ray, 
the relative velocity between critical and non-critical particles asymptotes 
the speed of light.}
\end{figure}

We would like to propose an intuitive physical explanation for the 
particle acceleration.
A black hole is defined by a region from which no light ray can escape. 
Since the event horizon is the boundary of the black hole 
region, it must be a null hypersurface on which the only possible 
causal curves are null geodesics. So we can interpret that the velocity 
of a critical particle, which asymptotes the event horizon in 
infinite proper time,
also approaches the speed of light. In fact, we can show that the
relative velocity of the critical particle 
with respect to a non-critical free-falling particle 
approaches the speed of light. This is the intuitive explanation why 
the CM energy of particle collision between critical and 
non-critical particles can be arbitrarily 
large near the horizon~\cite{Jacobson:2009zg}.
Figure~\ref{fg:conformal_diagram} schematically shows the 
trajectories of the two colliding particles in 
the spacetime diagram.

We can interpret it in a slightly different way.
No particle or no light ray can escape 
from behind the event horizon of the black hole.  
This can be understood that the velocity of a 
free-falling particle reaches the speed of light
with respect to the distant observer. 
It should be noted, however, that the relative velocity of 
distant two bodies does not have any definite physical meaning 
in general relativity.  
So, if it is possible for a particle to stay at a constant radius $r$
near the horizon, which is analogous to
a static observer in the $tr$ plane, 
the relative velocity of such a particle with respect to the
non-critical free-falling particle approaches the speed of light in 
the limit of the spacetime point to the horizon.
A similar discussion in terms of the velocity relative to the 
zero angular momentum observer (ZAMO) is found in~\cite{Zaslavskii:2011dz}.

Here it would be helpful to study 
the difference in the behaviour of the particles orbiting
near the horizon between a Schwarzschild black hole and a
Kerr black hole and see how our explanation works.
We focus on massive particles in the equatorial plane
for clarity.
{}For the Schwarzschild black hole, where the event horizon is located at
$r_{H}=2M$, any particle cannot stay in the vicinity of the
horizon.
This can be seen in terms of circular orbits, which can be located 
by $V(r)=V'(r)=0$. Using Eq.~(\ref{eq:effective_potential}) with $a=0$, 
we find two roots
\begin{equation}
 r_{\pm}=\frac{L^{2}\pm L\sqrt{L^{2}-12m^{2}M^{2}}}{2 m^{2}M},
\end{equation}
if $|L|\ge 2\sqrt{3}m M$. The energy of the particle can be 
determined by the condition $V(r_{\pm})=0$. 
$r_{+}$ and $r_{-}$ correspond to stable and unstable circular
orbits, respectively. We can see that $3M<r_{-}\le 6M$ and
$6M\le r_{+}<\infty $ for $2\sqrt{3}mM\le |L| $,
where the 
equalities $r_{+}=r_{-}=6M$ hold only for $|L|=2\sqrt{3}mM$
and $r=6M$ is called the innermost stable circular orbit (ISCO).  
The reason why high energy particle collision does not occur 
in the Schwarzschild spacetime is that there is no 
circular orbit in the vicinity of the horizon, whether stable or not. 
Any two particles near the horizon plunge into 
the horizon with the velocity of light with respect to a static 
observer and the relative velocity between the two particles
cannot be so large.
It is the innermost circular orbit that determines the possibility of 
high energy particle collision in the vicinity of the horizon, although 
the ISCO is also very important on its own, 
which will be discussed later.

For the Kerr black hole, the explicit expression of the circular 
orbits is complicated but we can easily see whether there is 
the one in the vicinity of the horizon. 
Using Eq.~(\ref{eq:effective_potential}), we find
\begin{eqnarray}
V(r_{H})&=&-\frac{(r_{H}^{2}+a^{2})^{2}(E-\Omega_{H}L)^{2}}{2r_{H}^{4}}, \\
V'(r_{H})&=&\frac{(r_{H}^{2}+a^{2})^{2}(E-\Omega_{H}L)^{2}}{8r_{H}^{5}}\nonumber
 \\
&& -\frac{2r_{H}(r_{H}^{2}+a^{2})E(E-\Omega_{H}L)-(r_{H}-M)[m^{2}r_{H}^{2}+(L-aE)^{2}]}{r_{H}^{4}}.
\end{eqnarray}
We should also note that the effective potential is analytic with
respect to $r$ at around $r=r_{H}$. Then, we can easily see that there is no
circular orbit in the vicinity of the horizon $r=r_{H}$, whether it is
stable or not, unless
$E-\Omega_{H}L\simeq 0$ and $r_{H}\simeq M$, i.e., the particle is 
at least nearly critical and the black hole is at least 
nearly maximally rotating.

\begin{figure}[tbh]
\begin{center}
\includegraphics[width=0.45\linewidth]{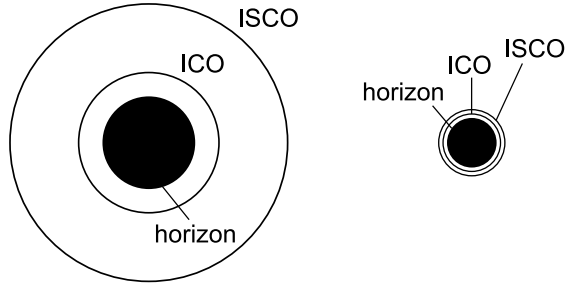}
\end{center}
\caption{\label{fig:isco_ico_fig_paper}
The locations of the innermost stable circular orbits (ISCO) and
the innermost circular orbits (ICO) on the equatorial plane are 
schematically shown. The left and right panels show the Schwarzschild 
black hole and a nearly maximally rotating Kerr black hole,
 respectively.}
\end{figure}

{}For a nearly maximally rotating black hole, 
the radii of both the innermost unstable circular orbit and ISCO
can be very close to 
that of the horizon. 
The situation is schematically shown in
Fig.~\ref{fig:isco_ico_fig_paper}.
As shown in~\cite{Bardeen:1972fi}, 
since all unstable circular orbits are located between the innermost 
unstable circular orbit and ISCO, the radii of unstable circular orbits
can also be very close to that of the horizon.
This implies that the geodesic particle
with $E-\Omega_{H}L\simeq 0$ 
can keep its radius constant in the vicinity of the horizon. 
Such a particle approximately satisfies 
the critical condition.
The relative velocity of a particle in the circular orbit near the horizon
with respect to the generic particle plunging into the horizon has a definite 
physical meaning if these two particles share the same 
spacetime point. This relative velocity approaches the 
speed of light in the limit to the horizon radius or in 
the maximal rotation limit, 
leading to the arbitrarily large CM energy for the 
collision of these two particles.

For a maximally rotating black hole, 
if we release a critical particle at rest at infinity, it approaches
the horizon in infinitely long proper time with infinitely many 
rotations because the horizon radius coincides with that of 
the maximum of the effective potential, although there is no
circular orbit for a massive particle on the event horizon~\cite{Harada:2010yv}.
The relative velocity of the critical particle coasting and approaching 
the horizon with respect to the generic particle plunging 
into the horizon has a definite 
physical meaning if these two particles share the same 
spacetime point and approaches the speed of light in the limit 
of the collision point to the horizon.

\section{Physical significance of particle acceleration}

\subsection{Criticisms and basic questions}
Just after the rediscovery by Ba\~nados et al.~\cite{Banados:2009pr},
several critical comments were given in Refs.~\cite{Berti:2009bk,Jacobson:2009zg}.
Here we pick up the following relevant ones: 
\begin{itemize}
 \item[(1)] There exists an upper bound on the 
spin parameter $a_* \lesssim 0.998$ for astrophysical black holes,
which is called Thorne's bound~\cite{Thorne:1974}. 
Then, the maximum value of $E_{cm}/(2m)$ is $\sim 9.49$
for the collision of particles of mass $m$ which are initially 
at rest at infinity~\cite{Harada:2010yv}.
 \item[(2)] The backreaction effect due to the absorption of a pair of the 
colliding particles of mass $m$ by a maximally rotating black hole
shifts the spin parameter $a_{*}$ from 1 to $1-2m/M$.
 \item[(3)] It needs arbitrarily long proper time for the 
critical particle to reach the horizon for the maximally 
rotating black hole. 
 \item[(4)] The radiation reaction becomes so large for the critical particle 
that the particle acceleration may be suppressed.
\end{itemize}
We add the following questions: 
\begin{itemize}
 \item[(5)] Is it possible to fine-tune the angular momentum in nature?
 \item[(6)] Is the collision of high CM energy restricted on the equatorial 
plane? 
 \item[(7)] How does the self-force of the particles affect the process?
\end{itemize}

Comment (3) has already been discussed in
Section~\ref{subsec:acceleration_time}. 
As for comment (1), Thorne's bound is based on the 
standard accretion disk model and hence at least 
model-dependent. See, e.g., Ref.~\cite{Abramowicz:1980, Sadowski:2011} 
for the possible
violation of this bound．

\subsection{Fine-tuning problem and the ISCO}

The analytic expression for the ISCO of the Kerr black hole
is given in Ref.~\cite{Bardeen:1972fi}.
The ISCO is taken as the inner edge of the accretion disk in the
standard accretion disk model.
A compact object which adiabatically inspirals around the black hole
undergoes a transition to a plunge phase at the ISCO in the limit 
of extreme mass ratio.
We can explicitly show that 
$r_{\rm ISCO}\to r_{H}$, $E_{\rm ISCO}\to m/\sqrt{3}$, $L_{\rm
ISCO}\to 2 m M /\sqrt{3}$ as $a_{*}\to 1$.
Noting 
$\Omega_{H}\to 1/(2M)$ as $a_{*}\to 1$, 
we find $E_{\rm ISCO}-\Omega_{H}L_{\rm ISCO}\to 0$.
In other words, a particle orbiting at the ISCO approaches the horizon and 
asymptotically satisfies the critical condition in the limit of
maximal rotation of the black hole. 
In fact, if particle 1 which orbits at the ISCO collides 
with particle 2 which is generic at the ISCO radius, we can find
\begin{equation}
\frac{E_{\rm cm}}{2m}\simeq \frac{\sqrt{2e_{2}-l_{2}}}{2^{1/6}3^{1/4}}
\frac{1}{\sqrt[6]{1-a_{*}^{2}}}
\label{eq:on-isco}
\end{equation}
for $a_{*}\simeq 1$, 
implying unboundedly high collision energy in the maximal rotation 
limit~\cite{Harada:2010yv}．
In view of the astrophysical significance of the ISCO, it turns out that 
the angular momentum of particles is naturally fine-tuned 
and $E_{\rm cm}$ can be very large for a rapidly rotating black hole.

\subsection{Non-equatorial orbits and collisions}

\begin{figure}[tbh]
\begin{center}
\includegraphics[width=0.35\linewidth]
{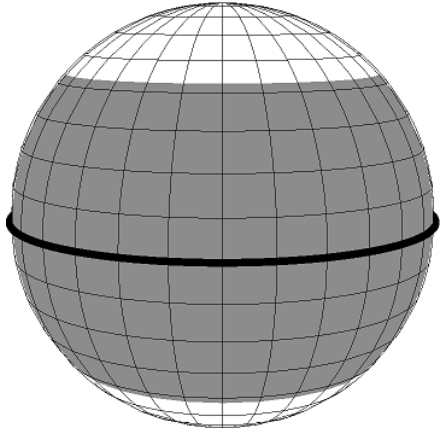}
\caption{\label{fg:collisionbelt} 
The grey region denotes the one where arbitrarily high energy 
collision can occur, while the white region denotes the one where 
cannot. The solid thick line denotes the equator with respect to the
 rotational axis.}
\end{center}
\end{figure}

As for question (5), let us consider 
particles which are not restricted in the equatorial plane.
If two general geodesic particles 
collide near the horizon, the CM energy is given 
by~\cite{Harada:2011xz} 
\begin{eqnarray}
E_{\rm cm}^{2}&=&
\frac{m_{1}^{2}r_{H}^{2}+{\cal Q}_{1}+(L_{1}-aE_{1})^{2}}
{r_{H}^{2}+a^{2}\cos^{2}\theta}
\frac{E_{2}-\Omega_{H}L_{2}}{E_{1}-\Omega_{H}L_{1}} 
+(1\leftrightarrow 2)+\cdots,
\end{eqnarray}
where ${\cal Q}$ is the Carter constant.
Therefore, we can see that $E_{\rm cm}$ is 
diverging if either of the two colliding particles
satisfies the critical condition.
As in the equatorial case, 
the critical particle can approach the horizon 
only if the black hole is maximally rotating. 
Furthermore, in this case, we can show that 
the polar angle $\theta$ must be 
in some range so that the latitude from the equator must be 
between $\pm 42.94^{\circ}$. 
Therefore, particle collision of arbitrarily high CM energy
occurs only on the region of latitude lower than $42.94^{\circ}$
and does not on the region of higher latitude.
See Fig.~\ref{fg:collisionbelt} for the schematic diagram of the 
high-velocity collision belt. 

\subsection{Effects of gravitational radiation reaction}

A critical particle rotates around the black hole infinitely 
many times with infinitely long proper time before reaching 
the horizon of the rotating black hole. 
One might think that the particle loses the considerable 
fraction of its energy and angular momentum through 
gravitational radiation and it results 
in the violation of the critical condition.
However, the radiation power $\dot{E}_{\rm GW}$ of the particle 
orbiting at the ISCO is strongly suppressed if the black hole 
is rapidly rotating and the detailed numerical calculation suggests that
it obeys the power law $\dot{E}_{\rm GW}\propto
(1-a_{*})^{\lambda}$ ($\lambda\simeq 0.317$) for 
$a_{*}\simeq 1$~\cite{Kesden:2011ma,Chrzanowski:1976}~\footnote{
Recently, a simple discussion has been proposed which suggests
that no gravitational wave is emitted from
a particle orbiting at the ISCO in the extremal limit
of the Kerr black hole~\cite{Hadar:2014dpa}.}.
Assuming that this power law holds and that the radiated energy 
due to the circularly orbiting particle is extracted from the kinetic
energy of the particle, we can follow the transition from the adiabatic 
inspiral phase to the plunge phase continuously and semi-analytically.
Within this framework, if the mass ratio of the particle to the black
hole
is sufficiently small and the black hole is nearly maximally rotating, 
very high CM energy is still attainable~\cite{Harada:2011pg}．

\subsection{Effects of self-gravity of the particles}
As for comment (2), it should be noted that the black hole can be 
efficiently spun up by continuous mass accretion. 
On the other hand, comment (2) suggests that 
the gravitational field generated by the 
particles may affect the process. The rest mass 
as well as the energy of each particle 
are assumed to be sufficiently small compared to 
the gravitational mass of the central black hole.   
It is expected, however, that the gravitational field 
generated by the two colliding particles cannot be neglected 
if the CM energy of the collision is comparable with the 
gravitational energy of the black hole.

Since the treatment of this problem is very difficult because of 
lower symmetry of the system, we will try to learn a lesson
from an analogous system with higher symmetry.  
For this reason, we replace the system of particles orbiting 
an axisymmetric rotating black hole 
with the system of electrically charged dynamical spherical shells 
around a static spherically symmetric electrically charged 
black hole,
which is given by the Reissner-Nordstr\"om black hole.
This is one of the examples where a very good analogy holds 
between the Kerr back hole and the Reissner-Nordstr\"om black hole.
Although the latter is apparently very different from 
the original system, it is very analogous to the original system as
the CM energy of the two spherical shells colliding in the vicinity of
the horizon can be arbitrarily large if the gravitational and
electromagnetic fields generated by each shell are 
neglected. See also~\cite{Zaslavskii2010_charged} for the collision of
two charged particles radially moving around a static 
spherically symmetric electrically charged black hole.
The advantage of the electrically charged black hole-shell system 
is that we can fully exactly take 
the fields generated by the shells into account.
It turns out that the CM energy $E_{\rm cm}$ of the two colliding
shells of equal proper mass 
in the vicinity but outside of the horizon is bounded as 
follows~\cite{Kimura:2010qy}: 
\begin{equation}
E_{\rm cm}
\lesssim 2^{1/4} M^{1/4} \mu^{3/4} , 
\end{equation}
where $\mu$ is the proper mass of each shell. Although the ratio 
$E_{\rm cm}/\mu$ can be very large if $M \gg \mu$,  
the boundedness of $E_{\rm cm}$ for the
finite values of $M$ and $\mu$ is important.

\section{Towards astrophysical black holes}
\subsection{Observability of high-energy particles}

The high energy collision of particles in the vicinity of the 
horizon can produce high energy and/or superheavy particles.
Can these particles be observed by a distant observer?
If such particles are to be emitted to infinity, extra energy 
gain is necessary as seen from simple energetics. 
In this context, it is known that the rotational energy of the 
black hole can be extracted. For this effect, the existence of 
the so-called ergoregion near the horizon of the rotating black hole, 
where the energy of the particle $E$
can be negative, plays an important role.

\begin{figure}[tbh]
\begin{center}
\includegraphics[width=0.85\linewidth]{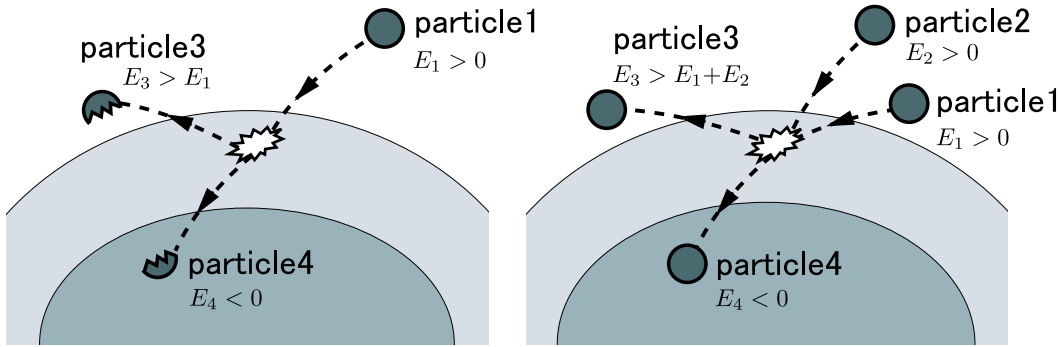}
\end{center}
\caption{\label{fg:penrose_processes}
The left and right panels are the schematic figures of the 
original Penrose process and the collisional Penrose process, 
respectively.
The dark and light shaded regions in each panel denote the black hole 
region and the ergoregion, respectively. 
}
\end{figure}

For a Penrose process~\cite{Penrose:1969}, 
the most representative energy extraction
process, particle 1 is released from infinity, disintegrates into 
particles 3 and 4 in the ergoregion and particle 3 escapes to infinity.  
If $E_{4}$ is negative, we have $E_{3}=E_{1}-E_{4}>E_{1}$, i.e., net 
positive energy gain. Since we are interested in 
the collision of two particles,  
let us instead consider two incident particles 1 and 2. Also in this case, 
if $E_{4}$ is negative, we have $E_{3}=E_{1}+E_{2}-E_{4}>E_{1}+E_{2}$, 
i.e. net positive energy gain. This process is called a collisional
Penrose process~\cite{Piran_etal_1975}.
In this process, we do not need to consider any artificial
disintegration process in the ergoregion, which is necessary in 
the original Penrose process. See Fig.~\ref{fg:penrose_processes} for the
schematic figures of both processes.

However, the requirement that particle 1 is a critical particle
and particle 3 produced in the vicinity of the horizon 
escapes to infinity turns out to be a very strong restriction.  
It can be shown that the energy of the product particle 3 can be at most
218.6 \% of that of incident particle
1~\cite{Bejger:2012yb,Harada:2012ap}.
The energy extraction efficiency $E_{3}/(E_{1}+E_{2})$ is at most 
137.2 \% for the inverse Compton scattering process, which 
is relatively more efficient among several important physical processes.

This does not necessarily mean that the high energy particle
collision is unobservable in principle. In fact,  
the observational effects from high-energy collision of dark matter 
particles surrounding a rapidly rotating black hole have been 
calculated~\cite{Banados:2010kn,Williams:2011uz}.
It is suggested that 
there can appear signature in the spectrum of gamma
ray emission of high energy collision of dark matter particles 
and subsequent pair annihilation around a black hole and 
this signature can be distinguished by the observation of the Fermi 
satellite~\cite{Cannoni:2012rv}.
However, consistently with the energy upper limit argument, 
it has been shown~\cite{McWilliams:2012nx} that the flux directly
emitted from the conventional Ba\~{n}ados-Silk-West process is
unmeasurably small because of strong redshift as well as greatly
diminished escape fraction. This point has been subsequently
acknowledged and the potential indirect observability 
has been discussed by other authors~\cite{Gariel:2014ara}.

\subsection{Effects of magnetic fields}
It is believed that there are strong magnetic fields 
around astrophysical black holes. The magnetic flux density 
is estimated to $\sim 10^{4}$ Gauss
around supermassive black holes and 
to $\sim 10^{8}$ Gauss around stellar mass black holes.
These magnetic fields are not so strong as to deform the 
black hole itself but strong enough to greatly affect the orbits of charged 
particles. 
The non-dimensional ratio $b$ of the Lorentz force to the 
gravitational force is estimated as
\begin{equation}
 b =\frac{qBGM}{m c^{4}}
\sim 10^{11}\left(\frac{q}{e}\right)
\left(\frac{m}{m_{e}}\right)^{-1}\left(\frac{B}{10^{8}\mbox{Gauss}}\right)\left(\frac{M}{10M\odot}\right),
\end{equation}
where $B$, $q$, $e$, $m_{e}$ are the magnetic flux density, the charge
of the particle, the elementary charge and the electron mass,
respectively,
and $G$ and $c$ are restored.
If the Lorentz force acts in the direction opposite to the gravitational
force, the charged particle has a smaller ISCO radius 
and a higher ISCO velocity  
than the neutral particle. This can be viewed as an indirect
acceleration of charged particles by the magnetic field, if we consider 
that a charged particle gradually shifts the radius of its circular
orbit towards the ISCO for the charged particle by radiation reaction.  
For a Schwarzschild black hole immersed in a uniform magnetic field, 
the CM energy for the collision of particle 1 which is a charged particle
orbiting at the ISCO against particle 2 which is a 
radially falling neutral particle
is~\cite{Frolov:2011ea}
\begin{equation}
E_{\rm cm}\simeq 1.74 b^{1/4}
m.
\end{equation}
For a Kerr black hole immersed in a uniform magnetic field, 
we can find~\cite{Igata:2012js}
\begin{equation}
E_{\rm cm}(b)\simeq \frac{\sqrt{b}}{3^{1/4}}E_{\rm cm}(0),
\label{eq:ecm_magnetic}
\end{equation}
where $a_{*} \simeq 1$ and  $b \gg 1$ are assumed 
and $E_{\rm cm}(0)$ is the CM energy in the absence of 
magnetic field, which is given by Eq.~(\ref{eq:on-isco})
for the collision of a particle orbiting at the ISCO.
Therefore, for the Kerr black hole, the magnetic field is expected to 
enhance the acceleration of charged particles 
to $\sim 10^{4}-10^{5}$ times higher than 
the value for the absence of the magnetic field.  
It will be interesting to study the acceleration of charged particles 
around a Kerr black hole with more realistic configuration of 
the magnetic field.

\section{Generalisations}
So far, we have focused on the collision of geodesic 
particles around a Kerr black hole.
In this section, we briefly review a variety of generalisations.

\subsection{High energy particle collision with bounded physical quantities}
In the process proposed by 
Ba\~nados, Silk and West~\cite{Banados:2009pr},
it is important that the CM energy $E_{\rm cm}$ can be 
arbitrarily large in the limit to the horizon, even though
the conserved quantities of the test particles with respect to a distant
static observer are finite.
We can separately discuss the condition for the divergence of $E_{\rm cm}$ 
leaving aside whether or not the colliding particles can reach
the horizon by any physical process,  
e.g., geodesic motion from the
distant region or continuous energy loss due to radiation.
We should refer to several works 
from this point of view.
Piran and Shaham~\cite{Piran_Shaham_1977_upper_bounds} discuss
that $E_{\rm cm}$ can be unbounded for the collision of ingoing particle
and outgoing particle with finite conserved quantities 
in the vicinity of the horizon
even in the
non-extremal Kerr black hole, although such a collision is not
physically well motivated. 
Grib and Pavlov~\cite{Grib:2010xj}
assume a near-critical particle in the vicinity of the
horizon which is inside the barrier of the effective potential. 
Although such a particle cannot reach the vicinity of the 
horizon from a distant region through any geodesic motion,
they invoke multiple scatterings.
They showed that such a near-critical particle can collide with 
a non-critical falling particle with unbounded CM energy
even around a non-extremal Kerr black hole. 
For these collisions, 
no physically realistic processes are known to give 
particles such special initial conditions.

\subsection{High energy particle collision in non-Kerr black holes}
As we have seen, Kerr black holes act as the accelerators of neutral particles.
In fact, the Kerr-Newmann family of black holes~\cite{Wei:2010vca,Liu:2011wv},
accelerating and rotating black holes~\cite{Yao:2011ai}
and Sen black holes~\cite{Wei:2010gq}
are shown to accelerate neutral particles to arbitrarily high
energy in the sense described in Section II.
These are the examples of extremal rotating black holes
which act as particle accelerators to unboundedly high energy 
if radiative reactions and self-force effects are neglected.

The Reissner-Nordstr\"om black holes are
inefficient in accelerating neutral particles
but can act as the accelerators of charged particles to unboundedly 
high energy~\cite{Zaslavskii2010_charged}.
It is also the case for 
general rotating and charged black holes~\cite{Zhu:2011ae}.
These are the examples of extremal charged black holes as 
the accelerators of charged particles to unboundedly high energy.

 Zaslavskii discussed high energy particle collision around 
``dirty'' black holes, which a certain class of stationary and 
axisymmetric black holes, including not only Kerr black holes but also 
black holes surrounded by matter distribution and black holes in 
alternative theories of gravity.
He showed that the situation is similar to the case of the 
Kerr black holes.
Irrespective of the explicit functional form of the metric, 
extremal dirty black holes are shown to accelerate neutral 
particles~\cite{Zaslavskii2010_rotating}.

Recently, the particle acceleration 
scenario has been extended to higher dimensions.
The Myers-Perry black holes, which are the higher-dimensional 
counterpart of the Kerr black holes, are shown to act as the accelerators
of neutral particles~\cite{Abdujabbarov:2013qka,Tsukamoto:2013dna}. 
The universality of particle collision of unbounded CM energy in the 
vicinity of the horizon of extremal black holes suggests a tight link 
to the field instability universally seen on the horizon of the 
extremal black holes shown by
Aretakis~\cite{Aretakis:2011ha,Aretakis_2011,Aretakis:2011gz,Aretakis:2012ei,Aretakis:2012bm,Aretakis:2013dpa} 
and others~\cite{Murata:2012ct,Murata:2013daa}.

\subsection{High energy particle collision in non black hole spacetimes}

At the event horizon of a black hole,
the infalling velocity of the free-falling particle might be understood 
as the speed of light with respect to a distant static observer.
We can interpret it as that 
the gravitational potential for such a particle is 
infinitely deep at the event horizon and hence photons emitted from 
such a particle are infinitely redshifted.

We should note that gravitational redshift can be very large
even in non black hole spacetimes. Photons emitted from a
particle in a compact region to infinity can be strongly redshifted
if the gravitational potential is very deep there.
We can also expect that high energy particle collision occurs 
in such a region~\cite{Patil:2012fu, Nakao:2013uj}.
In the absence of an event horizon, we can naturally consider
collisions between ingoing and outgoing particles in such a 
high redshift region. In fact, it has been shown 
that the CM energy can be very high for such particle collisions 
in several non black hole
spacetimes~\cite{Patil:2011ya,Patil:2011yb,Patil:2011aa,Patil:2012fu,Patil:2011uf}.
The efficiency and visibility of the high energy particle collisions 
around super-spinning near-extremal Kerr geometry are 
discussed in detail in~\cite{Stuchlik:2012zza,Stuchlik:2013yca}. 
This is in contrast to the case of black holes, where
the collision between ingoing and outgoing particles 
in the vicinity of the event horizon is not physically 
well motivated since the event horizon is a one-way membrane.
The non black hole scenario has a few possible 
advantages over the black hole one. 
It has been shown that in the system of charged spherical shells
the gamma factor between two colliding shells can be arbitrarily large 
even if the effect of self-gravity is fully taken into 
account in the context of the non black hole scenario~\cite{Patil:2011uf}.

\section{Conclusion}

We can expect that high energy particle collision occurs in the vicinity
of the horizon of a rapidly rotating black hole as a result of the 
particle acceleration in rather general
situations. This phenomenon is well founded on the properties of 
the geometry and the geodesic orbits of the extremal and near-extremal 
Kerr black holes.
On the other hand, the acceleration to infinitely high energy 
is unphysical and there should exist an upper bound on the CM energy
due to the finiteness of the acceleration time and 
probably due to the self-force of the colliding particles.
Although we do not identify this particle acceleration mechanism 
with the direct acceleration mechanism of observed cosmic rays,
it can imprint some indirectly observable signatures on the spectra 
and/or light curves of
cosmic rays, electromagnetic waves, neutrinos and gravitational waves.  

We have convincing evidence that the particle acceleration 
to arbitrarily high energy is one of the universal basic properties 
of extremal black holes not only in astrophysics but also in 
more general context. Here we have seen some of the simplest 
examples. 
Moreover, it should be noted that the
particle acceleration is seen not only in the vicinity of the event horizon of black
holes but also in the deep gravitational potential well in non black hole spacetimes.

\acknowledgments

The authors thank 
V.~Frolov, T.~Igata, P.S.~Joshi, T.~Kokubu, U.~Miyamoto, 
K.-I. Nakao, H.~Nemoto,  M.~Patil, 
J.~Silk, H.~Tagoshi, N.~Tsukamoto and O.B.~Zaslavskii
for fruitful discussion.
T.H. was partially
supported by the Grant-in-Aid No. 26400282 for Scientific
Research Fund of the Ministry of Education, Culture,
Sports, Science and Technology, Japan. 
M.K. is supported by a grant for research abroad from JSPS.

\end{document}